\documentclass[aps,twocolumn,superscriptaddress,showpacs,pre,floatfix]{revtex4-1}
\usepackage{graphicx,color}

\begin{document}

\title{Neutrinos from pion decay at rest  to probe proton strangeness in an underground lab}

\author{G. Pagliaroli}\email{giulia.pagliaroli@lngs.infn.it}
\affiliation{INFN, Laboratori Nazionali del Gran Sasso, Assergi (AQ), Italy}
\author{C. Lujan-Peschard}\email{carolup@fisica.ugto.mx}
\affiliation{INFN, Laboratori Nazionali del Gran Sasso, Assergi (AQ), Italy}
\affiliation{Departamento de Fisica, DCeI, Universidad de Guanajuato,  Le\'on, Guanajuato, M\'exico}
\author{M. Mitra}\email{manimala.mitra@durham.ac.uk}
\affiliation{INFN, Laboratori Nazionali del Gran Sasso, Assergi (AQ),  Italy}
\author{F. Vissani}\email{francesco.vissani@lngs.infn.it}
\affiliation{INFN, Laboratori Nazionali del Gran Sasso, Assergi (AQ),  Italy}
\affiliation{{Gran Sasso Science Institute (INFN), L'Aquila, Italy}}

\begin{abstract}
The study of the neutral current elastic scattering of neutrinos on protons at lower energies can be used as a compelling probe 
to improve our knowledge of the strangeness of the proton. 
We consider a neutrino beam generated from pion  decay  at rest, as 
provided by a cyclotron or a spallation neutron source and a    
1 kton scintillating detector with a potential similar to the Borexino detector.
Despite several backgrounds from solar and radioactive  sources it is possible to estimate two 
optimal energy windows for the analysis, one between \mbox{$0.65-1.1$} 
MeV and another between \mbox{$1.73-2.2$} MeV. The expected
number of neutral current events in these two regions, for an exposure
of 1 year,  is enough to obtain an error on the strange axial-charge 
10 times smaller than available at present.

\end{abstract}

\pacs{25.30.Pt; neutrino scattering. 14.20.Dh; properties of protons.}
\maketitle
\paragraph*{\bf Introduction:}
The neutrino-proton elastic scattering was proposed by Weinberg
\cite{weinberg} as a tool to investigate the neutral currents (NC).
 However, this reaction depends upon  `proton strangeness'--i.e., the
 strange-quark contribution to the axial form-factor of the proton--
a quantity that at present is not reliably known. Several theoretical \cite{theo} as well experimental investigations of 
neutrino-proton elastic scattering
\cite{ahrens,tayloe,mini} tried to probe this quantity. 
The best experimental result \cite{ahrens} found a value for the proton strangeness  
similar to the one predicted in \cite{theo}, but the 90\% C.L.\ range is compatible with zero. 

Proton strangeness matters in many contexts, e.g.: 
for the measurement of supernova neutrinos \cite{farr},  
for the couplings of certain dark matter candidates to nucleons \cite{ellis+karliner},
for the polarized parton distribution functions, e.g., \cite{emc}.
However, the flavor singlet term measured by deep inelastic 
scattering of charged leptons includes 
an anomalous gluonic contribution \cite{gluons}; 
unless one postulates that this contribution is zero \cite{abm}, a considerable amount of 
additional labor is needed to extract proton strangeness \cite{df}.

By contrast, a measurement of elastic neutrino-proton scattering probes 
this quantity directly. 
The low energy regime of this reaction offers important
advantages, as we will discuss in the following. 
{Until now, only the LSND experiment 
exploited low energy neutrinos
with the aim to measure proton strangeness \cite{hope1,hope2}. 
However the cosmic-ray related background was too high and   
made the result not competitive \cite{tayloe}. 
To overcome these limitations, we need to go in an underground site 
and we can exploit the recent developments of neutrino detectors.}
The Borexino detector, in particular, 
has reached an unprecedented low energy threshold for the detection 
of solar neutrinos \cite{BorexPRL}. 
This implies the possibility to
use highly pure scintillation detectors to measure protons with 
kinetic energies as low as 1.3 MeV. 
Thus, these technological achievements pave the way for new 
approaches to measure precisely the elastic scattering of neutrinos onto protons.

We will consider an artificial neutrino beam produced by pion decay at rest, 
as in the cyclotrons described in the DAE$\delta$ALUS proposal
\cite{daedalus,sim-daedalus} or 
in the SNS facility \cite{SajjadAthar:2005ke} (recall
that the  neutrino spectra 
from pion decay are  very well known, and have already proved being
useful for measurements of other cross sections \cite{karmen1st}). 
The  events are observed by a scintillation detector 
located   at $100$ m from the source and  
{hosted in an underground site}   
to have performances similar to Borexino \cite{BorexPRL}.
We explore the potential of this experimental setup and determine 
which sensitivity  can be reached  with 1 kton$\times$year of exposure.  
We discuss the dependence of the sensitivity on the experimental
features, showing how to rescale our findings for different
configurations; i.e., by varying the detector mass, the distance to the source, the time of data taking, etc.. 


\paragraph*{\bf Advantages of Low Energies:}

The amplitude of the NC transition 
$\nu(k) +p(p)\to \nu(k')+ p(p')$ is described by 
a matrix element of current-current type:
\begin{eqnarray}
\mathcal{M}=-\frac{G_F}{\sqrt{2}}\ \bar{u}_\nu' \gamma^a P_L u_\nu\   
 \bar{u}_p' \Gamma_a(q)  u_p  
\end{eqnarray}
where $G_F$ is the Fermi constant, $P_L\equiv(1-\gamma_5)/2$, $\bar{u}$ and $u$ are the spinors
of the outgoing and incoming particles, respectively. 
The vertex of the hadronic current $\Gamma_a(q)$ is,
\begin{eqnarray} \Gamma_a(q)=\gamma_a F_{1}(Q^2)  +i\frac{\sigma_{ab}q^b}{2 m_p} F_{2}(Q^2) 
 -  \gamma_a \gamma_5 G_{\mbox{\tiny A}}(Q^2)
\end{eqnarray}
where $q=p'-p$ is the transfered
momentum, $Q^2=-q^2$, $m_p$ is the proton mass and
$\sigma_{ab}=i[\gamma_a,\gamma_b]/2$. 
The vector form factors $F_{1,2}(Q^2)$  can be obtained by applying the 
conserved vector current hypothesis 
neglecting the strange vector form factors \cite{weinberg}.
The most important quantity for us is
the axial form factor $G_{\mbox{\tiny A}}(Q^2)$. 
This can be  connected to the 
analogous term in the charged current interactions, but only
up to the SU(2) isosinglet term, 
due to the strange quark axial current.
The effect of this term has been  
described by the parameter $\eta$ \cite{ahrens} :
   \begin{equation} \label{ax}
   G_{\mbox{\tiny A}}(Q^2)=\frac{g_{\mbox{\tiny A}}(0)\times  (1+\eta)}{2\ (1+Q^2/M_{\mbox{\tiny A}}^2)^2} 
   \end{equation}
where we adopt the customary dipole parameterization, 
$M_{\mbox{\tiny A}}$ is the axial mass and $g_{\mbox{\tiny
    A}}(0)=1.267$. 

Let~us  clarify the relation of $\eta$ with  a often used  quantity, $\Delta s$.
The axial part of the hadronic current sums the contribution of all quarks,
$\sum_{q} \bar{q}\gamma^\mu\gamma_5 q\, t_{3\mbox{\tiny L}}\!(q)$,
where $t_{3\mbox{\tiny L}}\!(q)=\pm 1/2$ \cite{weinberg}. Its   
matrix element on a proton state can be parameterized by the numbers 
$\Delta q$, defined  by
 $\langle p| \bar{q}\gamma^a \gamma^5 q|p \rangle\equiv 
 \bar{u}_p\gamma^a\gamma^5 u_p\cdot \Delta q$. Thus at tree level  
$ G_{\mbox{\tiny A}}(0)=\Delta u/2-\Delta d/2-\Delta s/2$, that agrees
with Eq.~(\ref{ax}) setting  $g_{\mbox{\tiny A}}(0)\equiv \Delta u-\Delta d $ and  
$\eta\equiv -\Delta s/g_{\mbox{\tiny A}}(0)$; the
heavy quarks enter at higher orders and require to
 replace
$\Delta s\to \Delta s -\mathcal{P} (\Delta u+\Delta d+\Delta s)$ 
\cite{bass} with $\mathcal{P} \sim -0.02$ \cite{bass2}.
In the rest of this work, the term  `proton strangeness'  will always indicate  $\eta$.

The BNL 734 experiment \cite{ahrens} obtained 
\begin{equation}\label{aha}
\eta=0.12\pm 0.07
\end{equation}
that implies $\eta=0-0.25 \mbox{ at }90\%\mbox{ C.L.}$, compatible with zero.
Moreover, due to the use of neutrinos with energies around  1 GeV, the 
experimental search for $\eta$ at \cite{ahrens,mini} were
strongly entangled with the high $Q^2$ 
behavior of the axial form factor, that is to some extent uncertain.
Conversely, the impact of the `axial mass' on the cross section 
is greatly reduced if lower energies are considered.

\begin{figure}[t]
\centering{
\includegraphics[width=0.3\textwidth,
angle=0]{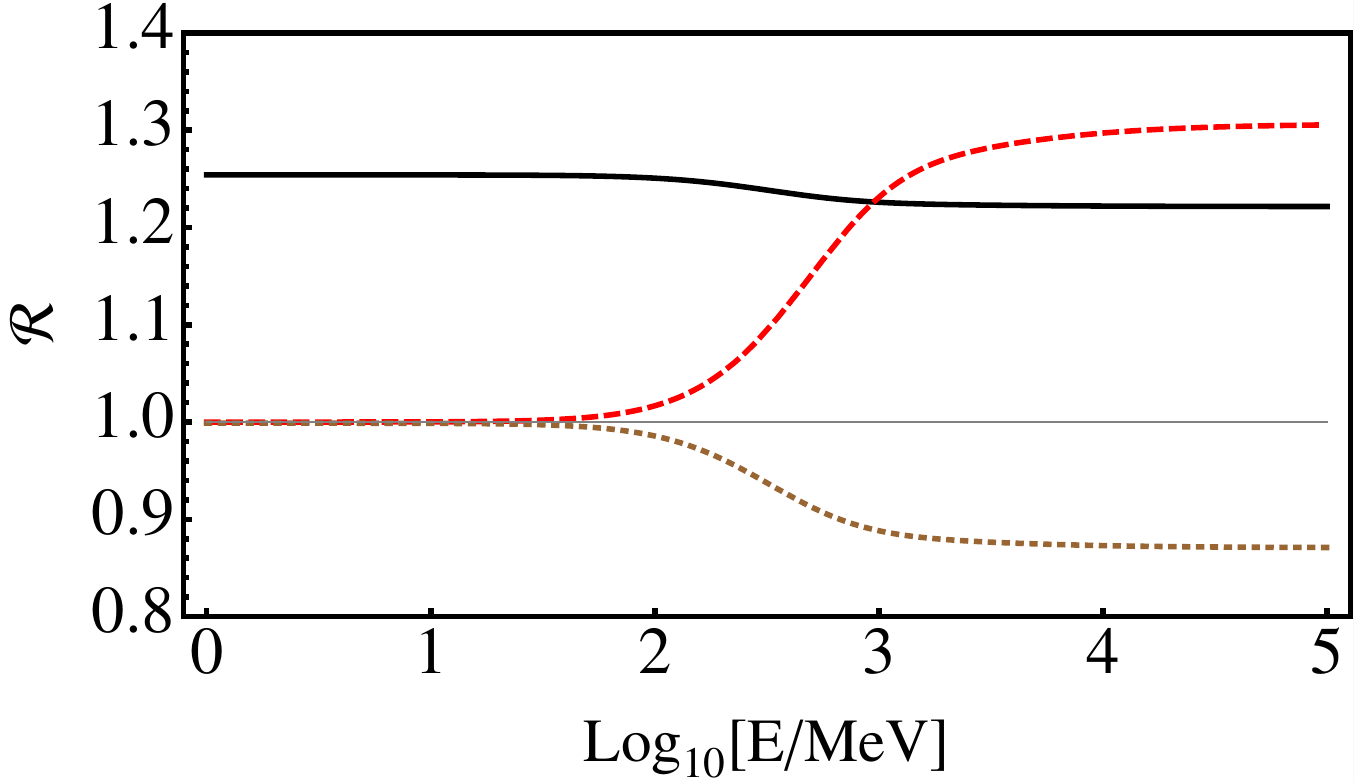}}
\caption{\em Variation of $\mathcal{R}(E_{\nu})$ with the
  neutrino/antineutrino energy $E_{\nu}$. 
The black solid line corresponds to 
$\eta=0.12$, while in the red dashed and brown dotted lines $\eta=0$. 
The red dashed line represents the effect 
of a relatively higher axial mass $M_A=1.2$ GeV in the numerator of
Eq.\ref{rat}. 
}
\label{xsec}
\end{figure}

This is illustrated in Fig.~\ref{xsec}, where we discuss how
the cross section changes, by displaying the ratio:
 \begin{equation}
 \mathcal{R}(E_{\nu})=\frac{\sigma_{\nu p}(E_{\nu})' + \sigma_{\bar{\nu} p}(E_{\nu})' }{\sigma_{\nu p}(E_{\nu}) + \sigma_{\bar{\nu} p}(E_{\nu})}.
\label{rat}
 \end{equation}
In the denominator, we put the reference cross section 
of \cite{ahrens} 
{(downloadable at \cite{xxsec})}   with $\eta=0$ and $M_A=1.032$
 GeV. In the numerator we modified it as follows:
1)~In the dotted curve, we omit the vector current contribution; 
at low energies,  the cross sections  does not change and 
$\mathcal{R}\to  1$. Note that $F_1(0)=1/2-2 \sin^2\theta_W\approx 0.04$, see also \cite{farr}.
2)~In the dashed curve, we set $\eta=0$ and use $M_A=1.2$ GeV;  
at GeV energies, this increases the axial form factor,  
the cross section and thus $\mathcal{R}$. 
3)~In the continuous curve, we use the central value $\eta=0.12$ of 
Eq.~(\ref{aha}) instead: The cross section and thus $\mathcal{R}$
increase at all energies.

In short, if we measure an increase of interactions at GeV energies, this can be attributed 
to a non-zero $\eta$ or to a larger value of  
$M_A$, see also \cite{ahrens}. 
Instead, for $E<\mbox{ few }100$ MeV the  cross section 
varies only with the axial form factor and more precisely with  
$ G_{\mbox{\tiny A}}(0)$: thus, a precise measurement of the
cross section probes  $\eta$.

\paragraph*{\bf Observable Spectra in Ultra-pure Scintillators:}

Let us consider an intense beam of neutrinos
produced from pion decay at rest according to the usual decay chains 
$\pi^+ \to \mu^+ +\nu_{\mu} \to e^+ +\nu_e+
\bar{\nu}_{\mu}+\nu_{\mu}$. The energy spectra of these 
neutrinos are very well-known and include a monochromatic line 
for the muonic neutrinos at $(m_\pi^2-m_\mu^2)/2m_\pi=29.8$ MeV and 
two continuous spectra with energy between $0<E_\nu<(m_\mu^2-m_e^2)/2m_\mu=52.8$ MeV. 
We assume $N_{\pi^+}=4.9 \times 10^{22}\ \text{pions}/\rm{yr}$  and thus, the same number of  $\nu_e$,
$\nu_\mu$ and $\bar\nu_\mu$. Such a pion production rate can be
provided, for example, by a cyclotron with $800$ MeV kinetic energy of
protons and with a peak power of $1$ MW \cite{daedalus}. 

We calculate the 
NC interactions of these $\nu_e$, $\nu_{\mu}$ and  $\bar{\nu}_{\mu}$ 
with the protons of a ultra-pure scintillation detector of 
mass 1 kton located at a distance of 100 meters and with the 
low energy performances as of Borexino detector \cite{BorexPRL}. 
The expected distribution of the elastic scattering events, $\nu (\bar{\nu})+ p
\to \nu (\bar{\nu})+ p $,  in the kinetic energy of the final state proton $T_p\equiv E'_p-m_p$ 
can be written as 
\begin{equation}
\frac{dN}{dT_p}=N_p \int_{E_{\mbox{\tiny min}}}\!\!\!\!\!\!\!\!\! dE_\nu \left[
\frac{d\sigma_{\nu}}{dT_p}(E_\nu,T_p)\ \Phi_{\nu}(E_\nu)   + (\nu\to\bar\nu)
 \right]
\end{equation}
where $N_p$ is the number of protons in the scintillator, $\sigma_{\nu, \bar{\nu}}$ are the 
cross sections and $\Phi_\nu$ and
$\Phi_{\bar{\nu}}$ are the total neutrino and antineutrino fluences (i.e., time integrated fluxes) 
differential in the neutrino energy $E_\nu$.
We assume the  chemical composition of Borexino, $C_9 H_{12}$; thus
the number of protons in $1$ kton is $N_p=6.02 \times 10^{31}$ (this
increases only by 20\% for the composition of LENA, $C_6 H_5 C_{n} H_{2n+1}$ 
with $n=12$ \cite{lena}).

The  proton kinetic energy $T_p$  and the 
minimum energy of the neutrino in the initial state are related as:
\begin{equation}\label{thth}
E_{\mbox{\tiny min}}=\frac{1}{2}\left[T_p+\sqrt{(2 m_p + T_p) T_p }\right].
\end{equation} 
The proton kinetic energy has to be converted into an `equivalent' detectable 
electron energy $E_d$, the {\em part} of $T_p$ that goes into 
scintillation light.  
This can be measured and parameterized \cite{ianni} and we adopt this
procedure. 
{Alternatively, one can use
the empirical Birks' formula 
$E_d=\int_0^{T_p} {dE}/(1+k_{\mbox{\tiny B}}\,  ({dE}/{dx}))$
\cite{birks,leo},
where $dE/dx$ is the stopping power of the proton, that depends 
on the chemical composition of the detector.} In our case,
the value $k_{\mbox{\tiny B}}\approx 0.011$ cm/MeV is consistent with
\cite{ianni} and agrees with \cite{v}.

For the purpose of the analysis we use a 
Gaussian energy resolution $\sigma(E_d)=50\ \mbox{keV} \sqrt{E_d/\mbox{MeV}}$
similar to Borexino  \cite{BorexPRL}.
Therefore, the detection
energy threshold $E_d^{th}=0.25$ MeV of Borexino
corresponds to a  
proton kinetic energy threshold 
of about $1.3$ MeV. From Eq.~(\ref{thth}), we see that we are
sensitive to the highest part of the neutrino energy spectra, $E_{\nu} \ge 25 $ MeV.

\begin{figure}[t]
\centering{
\includegraphics[width=0.45\textwidth,
angle=0]{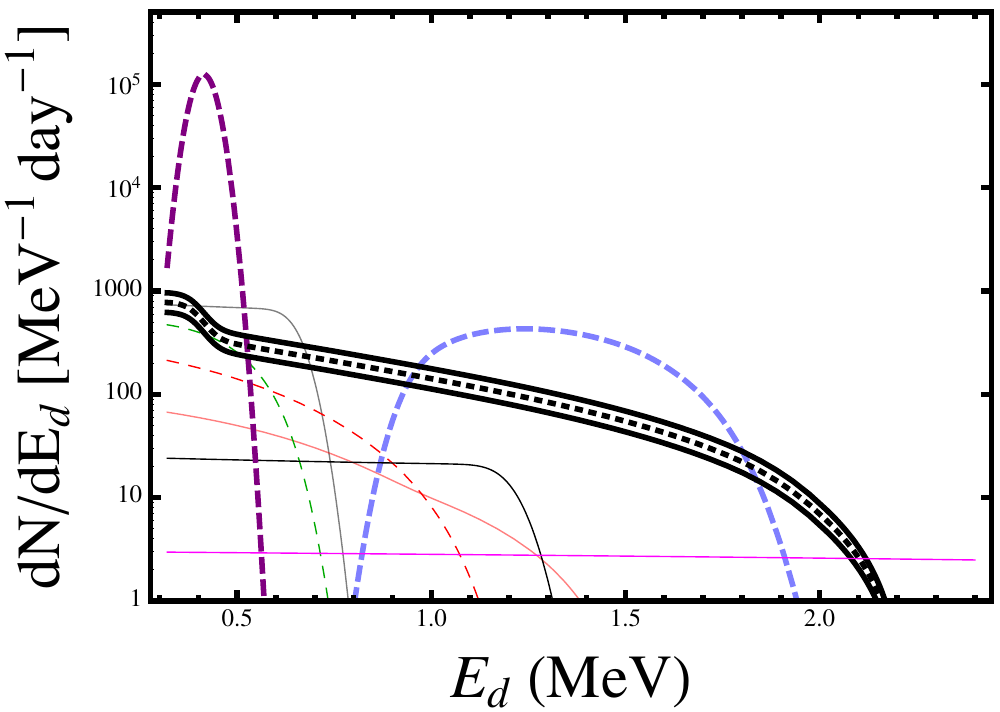}}
\caption{\em Expected event distribution in the detectable energy $E_d$ 
in 1 kton of scintillator as Borexino. The thick solid and dotted  black 
 lines  show  the event spectrum for neutral current elastic
 interactions of  neutrinos and antineutrinos with protons. The
 dashed blue line represents the cosmogenic $^{11}$C background. 
The magenta, pink, gray and black solid lines represent
 $^8B$, $CNO$, $^7Be$ and $PeP$  solar neutrino spectrum. The purple,
 red and green dashed lines represent the backgrounds due to  $^{210}Po$,
$^{210}Bi$ and $^{85}Kr$, respectively \cite{borex}. \label{fig2}}
\end{figure}

The differential spectrum $dN/dE_d$ expected for 1 year of 
exposure is shown in Fig.~\ref{fig2}. The signal due to NC $\nu p$
interactions is plotted using three black thick lines. 
They represent the events expected assuming 
$\eta=0.0, 0.12$ (best fit of \cite{ahrens}) and 0.25. 
From Fig.~\ref{fig2} we see that the energy window of interest for the detection of the elastic scattering interactions is $E_d=(0.25-2.2)$ MeV.
The number of signal events expected in this energy window 
and for one year of exposure is  considerable 

\begin{center}
\begin{tabular}{|c||c|c|c|}
\hline
strangeness $\eta$  & 0.00 & 0.12 & 0.25\\
\hline
signal events \hspace{0.5cm}& $0.89 \cdot 10^{5}$& $1.14 \cdot
10^{5}$ & $1.41 \cdot 10^{5}$\\
\hline
\end{tabular} 
\end{center} 
  
However, the same energy window will also contain
several solar as well as radioactive background events. In particular
solar neutrino events from $^{7}Be$, $CNO$, $PeP$ and $^{8}B$ are not negligible; we 
expect about $28.6$, $3.0$, $2.1$ and $0.5$ counts per day (cpd) in each 100 tons of scintillator respectively.
Moreover, the  backgrounds due to radioactive sources  as $^{210}Po$,
$^{210}Bi$ and $^{85}Kr$, as well as the cosmogenic $^{11}C$ have to
be considered. We assume that each 
100 tons of scintillator yield $10^3$ cpd from
$^{210}Po$, $25$ cpd from  both 
$^{85}Kr$ and $^{11}C$ as reported in \cite{borex}. For the $^{210}Bi$
we consider $15$ cpd consistent with the present Borexino level
\cite{aldo}. The spectra for these backgrounds are displayed in Fig.~\ref{fig2}.

Note incidentally that the Borexino detector is still  improving on
the background, and in the most recent runs, the ${}^{210}Po$
contaminant has already been halved and the $^{85}Kr$ decreased more
than three times \cite{aldo}. 
The contribution of the $^{11}C$ is particularly critical for our
purposes, and depends strongly on the depth of the underground site of
the detector \cite{galbiati}.  The value of this cosmogenic component is estimated to 
be 3 times smaller in LENA \cite{lena} and much less in SNO+ \cite{snop}.
However, in the present study we will adopt the conservative
background levels discussed above.

{Several meters of rock will moderate the fast neutrons that are subsequently absorbed \cite{bas}.}
The irreducible background is due to the neutrinos
from the cyclotron. In the region between $(0.25-2.2)$ MeV and for
each 100 tons of scintillator, 
we expect $0.12$ cpd due to elastic scattering on electrons, 
$0.1$ cpd due $\nu+^{12}\!C \to\nu+ p+ ^{11}\! B$, 
$0.5$ cpd due to neutron capture following  
$\nu+^{12}\! C \to \nu+ n+ ^{11}\! C$, 
$0.5$ cpd due to the protons related to 
$\nu_e+^{12}\!C \to p +e^- + ^{11}\!C$ \cite{Yoshida}. 
All these count rates are at most
at the level of $^8\!B$ background (magenta line in Fig.\ref{fig2}). 
Two of these reactions end with $^{11}C$ nuclei. The related increase of the 
cosmogenic component amounts to $2.5$ cpd per 100 ton.
Thus, these irreducible background processes have a minor role and can be safely neglected.

{\bf Sensitivity to proton strangeness $\eta$:}
From Fig.~\ref{fig2}, it is evident that background sources can pollute the
extraction of the signal. However, following \cite{vilante-aldo}, we proceed to 
extract the signal from the total number of observed events, measuring the 
background events and subtracting them. 

If we collect $N=S+B$ events in a time $T$ 
and $N_0=B\cdot T_0/T$ events in a time $T_0$, when the signal is absent,
we expect $S=N-N_0 \cdot T/T_0$ signal events.
Since both $N$ and $N_0$  are subject to Poissonian fluctuations, the 
fractional uncertainty caused by the statistical effects 
is given by:
\begin{equation}
\frac{\Delta S}{S}=\sqrt{\frac{1+\frac{B}{S} \left( 1+\frac{T}{T_0}\right)}{S}} \label{e}
\end{equation}
If observations are binned in energy, 
we define $2/\Delta S^2\equiv \partial^2( \chi^2)/\partial S^2$
using  $\chi^2=(S_i-\bar{S}_i)^2/\Delta S_i^2$ and $S_i =S\rho_i$,
with $\sum_i \rho_i=1$.
In the minimum of the $\chi^2$,  $S_i=\bar{S}_i$, 
$
\frac{\Delta S}{S} = \left\{ {\sum_{i=1}^n S_i/\left[1+ \frac{B_i}{S_i}\left( 1+\frac{T}{T_0}\right)\right] }\right\}^{-1/2}
$.
This is smaller than Eq.~(\ref{e})  where we set $S=\sum_{i} S_i$ and 
$B=\sum_{i} B_i$, unless 
the signal and the background have the same shape. 
For constant background-signal ratio, $B_i/S_i=$constant, 
both the results coincide.

From Fig.~\ref{fig2} it is evident  that there are two comparatively  
clean energy windows to extract the signal, 
one between $(0.65-1.1)$ MeV and another between $(1.73-2.2)$ MeV respectively.
Using $T=T_0=$1 year,  
we expect  $S_1=28900$ and $B_1=37200$  events
 from the first  energy window 
 and $S_2=1900$ and $B_2=2400$  events from the 
 latter one. 
Since we have $B_1/S_1\simeq B_2/S_2$ we can use
 Eq.~(\ref{e}) obtaining $\Delta S/S=1.1$ \%, dominated by the data in the first energy window. If we 
consider that  for a typical pulsed signal \cite{daedalus}, 
$T_0$ is 4 times larger than $T=1$, the error decreases to $\Delta S/S=0.9$\%; 
a similar improvement is obtained if we measure the background 
over the range $0.65<E_d<2.5$ MeV, increasing the statistic and thus 
reducing its error.

The NC cross section $\sigma$ at low energy 
(and therefore the signal event rate  $S$) is dominated by the axial form factor, 
scaling as $(1+\eta)^2$, as is clear from Eq.~(\ref{ax}).
So we can relate the signal sensitivity to the one on  $\eta$
through $\delta S=\Delta \sigma/\sigma=2 \Delta \eta/(1+\eta)$.
This means that the sensitivity $\Delta S/S=1.1$\% on the NC signal
implies that proton strangeness can be measured with an absolute error of 
$\Delta \eta=6\cdot 10^{-3}$ for $\eta=0.12$ 
(similar in size to the heavy quark corrections \cite{bass}) 
i.e., a $5\%$ fractional uncertainty. 
For all values of $\eta$ in the allowed experimental 
range, the error is more than 10 times smaller than Eq.~(\ref{aha}).

Finally, we show how to attain the same sensitivity with different experimental parameters. 
Let us write the uncertainty in $\eta$ as follows,
\begin{equation}
\Delta \eta = \frac{1+\eta}{2}  
\sqrt{\frac{1+\frac{1}{\varphi}\, \frac{B}{S} \left( 1+\tau\, \frac{T}{T_0}\right)}{\varepsilon\, \varphi\, S}} 
\label{sens1}      
\end{equation}  
Here   $\varepsilon=(M'/M) (T'/T)$ rescales the exposure, 
proportional to the mass $M'$ and to the time of data taking $T'$;
$\tau=(T'/T_0')/(T/T_0)$ rescales the fraction of data taking time 
that includes the signal;
$\varphi=(P'/P) (D/D')^2$ rescales the flux, 
accounting for the new power $P'$ and the  new 
distance $D'$.
Let us adopt the operational parameters of \cite{daedalus}, i.e.,
assume a distance from the cyclotron of $D'=700$ m, 
a power increased by a factor of 5 and $T'=T_0'/4=1$ year. The flux (and the signal) 
decreases by one order of magnitude, for it 
increases with the power but decreases as the distance squared. In order to compensate for this, 
we need a larger exposure. E.g., with a mass  of 50 kton as in the LENA proposal \cite{lena}
we obtain again $\Delta S/S=1.1$ \% after 1 year. 
Note that the selection of the optimal distance between source and
detector will require to minimize the beam related backgrounds.

\paragraph*{\bf Summary:}
In this work, we explored  the possibility to study proton strangeness using low energy neutrinos. 
We discussed the potential of exploiting a synergy between artificial neutrino beams from pion decay and
ultra-pure scintillating detectors. Incidentally, such complex will
allow us to  quantify precisely the response of underground neutrino
detectors, to search for sterile neutrinos and to
investigate many more issues.

We showed that, by using reasonable assumptions on the experimental
parameters, it is possible to identify two energy windows that 
allow to measure many tens of thousand of signal events. They imply a
statistical error on proton strangeness one order of magnitude smaller
than obtained by BNL 734, Eq.~(\ref{aha}). 
It is important to emphasize that the signal we discussed should be
collected above 0.6 MeV, 
and thus, it does not require the extreme low energy threshold 
reached by Borexino detector. 


\acknowledgments
\vskip-2mm
We thank  
A~Bacchetta,
G~Bruno, 
W~Fulgione,
G~Garvey, 
A~Ianni, 
M~Mannarelli,
A~Molinario,
A~Thomas,
F~Villante 
{and 
the Referees}  
for useful discussions.

\end{document}